# Orbital angular momentum conversion of optical field without spin state

Zhongsheng Man[1,2,3], Yudong Lyu[1], Zhidong Bai[1], Shuoshuo Zhang[1], Xiaoyu Li[1], Jinjian Li[1], Changjun Min[2,4], Fei Xing[1], Xiaolu Ge[1], Shenggui Fu[1], and Xiaocong Yuan[2,5]

[1]*School of Physics and Optoelectronic Engineering, Shandong University of Technology, Zibo 255000, China*

[2]*Nanophotonics Research Centre, Shenzhen University & Key Laboratory of Optoelectronic Devices and Systems of Ministry of Education and Guangdong Province, College of Optoelectronic Engineering, Shenzhen University, Shenzhen 518060, China*

[3]*zsman@sdut.edu.cn*

[4]*cjmin@szu.edu.cn*

[5]*xcyuan@szu.edu.cn*

**Abstract:** As one fundamental property of light, the orbital angular momentum (OAM) of photon has elicited widespread interest. Here, we theoretically demonstrate that the OAM conversion of light without any spin state can occur in homogeneous and isotropic medium when a specially tailored locally linearly polarized (STLLP) beam is strongly focused by a high numerical aperture (NA) objective lens. Through a high NA objective lens, the STLLP beams can generate identical twin foci with tunable distance between them controlled by input state of polarization. Such process admits partial OAM conversion from linear state to conjugate OAM states, giving rise to helical phases with opposite directions for each focus of the longitudinal component in the focal field.

**Keywords**：orbital angular momentum; spin angular momentum; optical vortex; state of polarization; optical singularity; vector beams; vector diffraction theory.

# 1 Introduction

It is well-known that a light beam can possess angular momentum (AM), in addition to linear momentum [1-11]. There are two categories of AMs including spin angular momentum (SAM) and orbital angular momentum (OAM). SAM is intrinsic and related to the vectorial nature of light, and it has two possible quantized values of $\pm\hbar$ depending on the handedness of circular polarization: $+\hbar$ per photon for a left-handed circularly polarized beam and $-\hbar$ per photon for a right-handed circularly polarized beam, where $\hbar$ is Planck's constant $h$ divided by $2\pi$ [1,2]. By contrary, OAM has both intrinsic and extrinsic terms, and the latter of which is coordinate dependent [5]. The intrinsic OAM, hereafter simply referred to as OAM, is related to the azimuthal dependence of optical phase. When a light beam possesses a vortex phase of $\exp(il\varphi)$, it can carry an optical OAM of $l\hbar$ per photon, where $l$ is topological charge, indicating the repeating rate of $2\pi$ phase shifts azimuthally along the beam cross-section [3-11]. Such a vortex beam exhibits a helical wave-front and possesses a phase singularity at the beam center, resulting in a doughnut-shaped intensity profile [12,13].

Since the discovery of light's OAM [3], optical vortices have provided insights into the fundamental properties of light and lead to abundant applications, including micromanipulation [14,15], optical communication [16-21], super-resolution imaging [22-24], quantum information processing [25,26], and others. Great successes have been achieved in the creation and manipulation of optical OAM. Conventionally, OAM beams may be generated in various ways like spiral phase plate [27-30], computer-generated holograms [31,32], sub-wavelength gratings [33]. These techniques rely on introducing a phase discontinuity in the wave-front to generate beams with desired OAM modes. Generally, it is believed that polarization and phase are two relatively independent degrees of freedom (DoFs) of light that show little interaction. Nevertheless, under specific conditions, the intrinsic optical DoF of polarization also enables the manipulation of optical OAM states via the procedure refer to as spin-orbital-conversion (SOC) [34-43]. Such a kind of SOC provides a direction connection between SAM (circularly polarized state) and OAM and allows for a broadband manipulation of OAS states, in contrast to the aforementioned modulators that are generally wavelength dependent. For the traditional SOC, the mapping from SAM is limited to circular polarization (CP). The most general state of polarization (SoP) is elliptical polarization (EP). CP is only one extreme case of an infinite set of EPs, and the other is linear polarization (LP). Most recently, the conversion of arbitrary SAM states (the general elliptically polarized states) into states with independent values of OAM is also achieved [44]. Such process requires the interaction of light with matter that is both optically inhomogeneous and anisotropic. As a counterpart, it has been proved that a direct transformation of the optical AM for the spin form to the orbital form can take place in homogeneous and isotropic medium provided that the input SoP is circularly polarized [35]. However, when the input optical field has not any spin states [every light vibration is linearly polarized for each photon in the beam cross-section], the achievement of such conversion is still a challenge in homogeneous and isotropic medium.

Here, we theoretically demonstrate that the OAM conversion of light without any spin state can indeed occur in homogeneous and isotropic medium when a specially tailored locally linearly polarized (STLLP) beam is strongly focused by a high numerical aperture (NA) objective lens. Based the vector diffraction theory, the analytical mode is built to calculate the three-dimensional electromagnetic field and Poynting vectors in the focal region of the proposed STLLP beam. Through a high NA objective lens, the STLLP beams can generate identical twin foci with tunable distance between the two controlled by input SoPs for all the transverse, longitudinal, and total fields. Such process provides partial conversion from linear polarization to conjugate OAM states, resulting in helical phases with opposite directions for each focus of the longitudinal component in the focal field.

## 2 Theory

The OAM conversion of light without any spin states in homogeneous and isotropic medium here is defined as the physical process by which the incident STLLP beam without any spin and orbital states can be partially transformed into two conjugated OAM states, resulting in two helical wave-front with opposite topological charges of −1 and 1 for the longitudinal component in the focal field, respectively. Figure 1(a) illustrates the concept of the OAM conversion. To realize such a physical process, there must be two elements: a high NA objective lens and specially-tailored locally-linear SoP for the input optical field. Polarization as an intrinsic optical DoF is one of the most salient features of light. For the well-known homogeneously polarized lights, their SoPs are spatially-invariant in the beam cross-section and can be described geometrically on the surface of a Poincaré sphere (PS) [45], as depicted in Fig. 1(b). Three parameters $s_1$, $s_2$, and $s_3$ describe the Stokes parameter of a point on the PS in the Cartesian coordinate system respectively, satifying $s_1^2 + s_2^2 + s_3^2 = 1$, and $2\phi$ and $2\theta$ denote the longitude and latitude angles of the point in the spherical coordinate system. The SoP at a given point ($2\phi$, $2\theta$) on the PS can be characterized by the unit vector **P** as follows [46]

$$\begin{aligned}\mathbf{P}(2\phi, 2\theta) &= \sin(\theta+\pi/4)\exp(-i\phi)\hat{\mathbf{e}}_r + \cos(\theta+\pi/4)\exp(i\phi)\hat{\mathbf{e}}_l \\ &= \frac{1}{\sqrt{2}}\left[\sin(\theta+\pi/4)\exp(-i\phi) + \cos(\theta+\pi/4)\exp(i\phi)\right]\hat{\mathbf{e}}_x + \\ &\quad \frac{i}{\sqrt{2}}\left[\sin(\theta+\pi/4)\exp(-i\phi) - \cos(\theta+\pi/4)\exp(i\phi)\right]\hat{\mathbf{e}}_y\end{aligned} \quad (1)$$

where {$\hat{\mathbf{e}}_r$, $\hat{\mathbf{e}}_l$} are right-handed (RH) and left-handed (LH) unite circular vectors, respectively, while {$\hat{\mathbf{e}}_x$, $\hat{\mathbf{e}}_y$} are horizontal and vertical unit linear vectors, respectively.

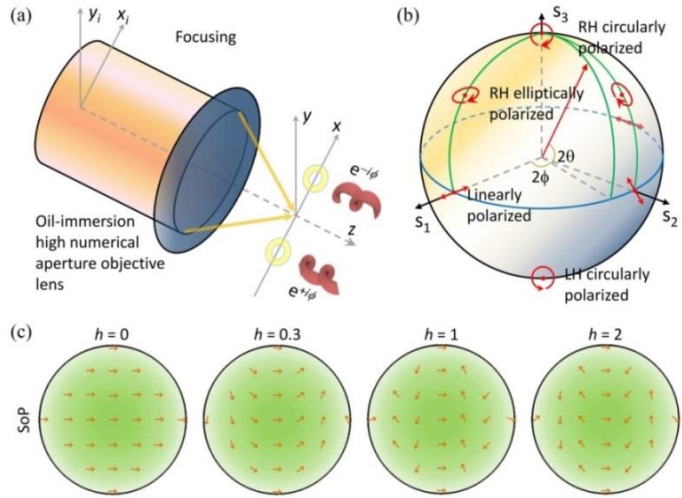

Figure 1: OAM conversion of optical field without any spin states achieved with the cooperation of a high NA objective lens and STLLP beam. (a) Schematic of a tightly focused system and coordinate system we followed in our calculations. The focal plane is located at $z = 0$. (b) PS representation of SoPs for homogeneously polarized lights. The poles represent circular polarization, the equator linear polarization and the intermediate points elliptical polarization. The northern and southern hemispheres stand for the right-handed and left-handed elliptical polarization. The polarization states at antipodal points are orthogonal. (c) Polarization distributions of four kinds of STLLP beams in terms of $h = 0$, 0.3, 1, and 2, respectively.

In additional to the simplest and most fundamental homogenous SoPs, a light wave admit spatially inhomogeneous SoPs, which is the so-called vector optical field (VOF) [46-56]. The SoPs of the most general EP beams can be described by a linear combination of a pair of orthogonal base vectors theoretically, so does the VOFs. For any pair of points on PS with the inverse symmetry with respect to the origin, the two SoPs can be serve as a pair of orthogonal base vectors, since $\langle \mathbf{P}_1(2\phi, 2\theta) | \mathbf{P}_2(2\phi + \pi, -2\theta) \rangle = 0$. When referring to VOF, its SoP may be represented by a unit vector $\mathbf{V}$ as follows [46-56]

$$\mathbf{V} = \frac{1}{\sqrt{2}} \{\exp(i\delta)\mathbf{P}_1 + \exp(-i\delta)\mathbf{P}_2\}. \tag{2}$$

Obviously, the SoP of VOF is a linear combination of orthogonal base vector vortices with opposite topological charge. Note that, the local SoPs in the beam cross-section of the VOF described by Eq. (2) are all in the circle of PS that is orthogonal to the connecting line between the two points $\{(2\phi, 2\theta), (2\phi + \pi, -2\theta)\}$. Hence, when the VOF is locally linearly polarized (LLP), $\mathbf{P}_1$ and $\mathbf{P}_2$ should satisfy $\mathbf{P}_1 = \hat{\mathbf{e}}_r$ and $\mathbf{P}_2 = \hat{\mathbf{e}}_l$, then we have

$$\mathbf{V}_{\mathrm{LLP}} = \frac{1}{\sqrt{2}} \{\exp(i\delta)\hat{\mathbf{e}}_r + \exp(-i\delta)\hat{\mathbf{e}}_l\} = \cos\delta\hat{\mathbf{e}}_x + \sin\delta\hat{\mathbf{e}}_y. \tag{3}$$

Apparently, the local SoP of the VOF described by Eq. (3) is linearly polarized because the orthogonal base vectors in term of $x$ and $y$ components are always in phase. The function $\delta$ can have arbitrary distributions in theory. For achieving the OAM conversion of light without any spin states in homogeneous and isotropic medium, $\delta$ should be given as follows

$$\delta = h_0 x = \frac{2\pi h}{r_0} r \cos\varphi, \tag{4}$$

where $h_0$ and $h$ are the horizontal indexes of polarization, $x$ is the Cartesian coordinate, $r$ and $\varphi$ are the polar radius and azimuthal angle, respectively, and $r_0$ is the radius of the VOF. Figure 1(c) depicts, respectively, the corresponding polarization distributions of four STLLP beams with $h = 0, 0.3, 1$, and $2$ when $r_0 = 1$ in Eqs. (3) and (4). For $h = 0$, it is LP beam and the local SoPs are *spatially-invariant*, which can be seen as one extreme case of STLLP beam, while for the other three, they are general STLLP beams and the local SoPs are *horizontally-variant*. Note that, the STLLP beams proposed here are much different from the previously reported VOFs, since no polarization singular points or lines can be found in the beam cross-section, which are new members of VOFs. There are geometric representations to describe the SoPs of VOFs, referring to higher-order or generalized PSs [57-59]. However, both of them are limited to describe VOFs with *azimuthally-variant* SoPs and cannot describe the SoPs of the proposed VOFs here, which is *horizontally* (*spatially*)-*variant*, since $\delta$ is a function of both azimuthal angle $\varphi$ and polar radius $r$, shown in Eq. (4). To study the OAM conversion mentioned above we consider the Laguerre-Gaussian ($LG_{l,p}$) laser modes [4] where $l$ and $p$ are, respectively, the numbers of interwined helices known as the topological charge and the additional concentric rings. Therefore, for a monochromatic paraxial STLLP-$LG_{l,p}$ beam, its electric field may be written as

$$\mathbf{E}_i \propto \left(\frac{r\sqrt{2}}{w_0}\right)^{|l|} \exp\left(-\frac{r^2}{w_0^2}\right) L_p^{|l|}\left(\frac{2r^2}{w_0^2}\right) \exp(il\varphi) \left[\cos\left(\frac{2\pi h}{r_0} r\cos\varphi\right)\hat{\mathbf{e}}_x + \sin\left(\frac{2\pi h}{r_0} r\cos\varphi\right)\hat{\mathbf{e}}_y\right], \quad (5)$$

where $L$ is the generalized Laguerre polynomials and $w_0$ is the radius of the beam waist.

The other key to achieve the OAM conversion mentioned above is the high NA objective lens. Strong focusing is required in many practical application ranging from microscopy to data storage as well as micro manipulation, and the vectorial properties of light change greatly after focusing [60,61]. Richards and Wolf originally considered the contribution of the input polarization to the focal field and built vectorical diffraction theory [60]. Here, we follow Richards and Wolf to calculate the electric field of an input field focused through an aplanatic high NA objective lens obeying the sine condition. When the incident polarized light is embodied by Eq. (5), its electric field in the vicinity of focus can be derived in the cylindrical coordinate system ($\rho$, $\phi$, $z$) as

$$\mathbf{E}(\rho,\phi,z)=\frac{-ikf}{2\pi}\int_0^{2\pi}\int_0^{\alpha}\sin\theta\sqrt{\cos\theta}l(\theta)\exp(il\varphi)\mathbf{M_e}e^{ik[-\rho\sin\theta\cos(\varphi-\phi)+z\cos\theta]}\mathrm{d}\varphi\mathrm{d}\theta, \qquad (6)$$

$$\mathbf{M_e} = \begin{bmatrix} \left[\sin\left(\varphi-\dfrac{2\pi h\sin\theta\cos\varphi}{\sin\alpha}\right)\sin\varphi+\cos\left(\dfrac{2\pi h\sin\theta\cos\varphi}{\sin\alpha}-\varphi\right)\cos\theta\cos\varphi\right]\hat{\mathbf{e}}_x \\ \left[\sin\left(\dfrac{2\pi h\sin\theta\cos\varphi}{\sin\alpha}-\varphi\right)\cos\varphi+\cos\left(\dfrac{2\pi h\sin\theta\cos\varphi}{\sin\alpha}-\varphi\right)\cos\theta\sin\varphi\right]\hat{\mathbf{e}}_y \\ \left[\cos\left(\dfrac{2\pi h\sin\theta\cos\varphi}{\sin\alpha}-\varphi\right)\cos\theta\right]\hat{\mathbf{e}}_z \end{bmatrix}, \qquad (7)$$

where $k$ and $f$ are the wave number in the image space and focal length of the focusing objective lens, respectively; $\theta$ and $\varphi$ denote, respectively, the tangential angle with respect to $z$ axis and the azimuthal angle with respective to $x_i$ axis; $\alpha = \text{arc sin}(NA/n)$, where NA is the numerical aperture and $n = 1$ is the index of refraction in the focal space; $\mathbf{M_e}$ is the electric polarization vector in the tightly focused field; and the function $l(\theta)$ describes the apodization function, which is given by

$$l(\theta)=\left(\sqrt{2}\beta\frac{\sin\theta}{\sin\alpha}\right)^{|l|}\exp\left[-\left(\beta\frac{\sin\theta}{\sin\alpha}\right)^2\right]L_p^{|l|}\left[2\left(\beta\frac{\sin\theta}{\sin\alpha}\right)^2\right], \qquad (8)$$

where $\beta$ is the ratio of the pupil radius [$r_0$] and the beam waist [$w_0$] that we take as 1 in our configurations.

## 3 Results and discussions

### 3.1 Splitting of input optical field into twin identical sub-optical-fields

For the STLLP-LG$_{0,0}$ beams, which is not carrying any SAM and OAM per photon [4], we simulated the total electric field intensity in the focal plane [NA = 0.95, $n$ = 1] with $h$ = 0, 0.3, 1, and 2, respectively. Obviously, the horizontal index $h$ affects the focal field strongly after tight focusing, shown in Fig. 2. To be specific, it is a single hot spot distribution with elliptical-shaped pattern for $h$ = 0 as seen in Fig. 2(a), and the long axis of which is in accordance with the direction of the input polarization [Fig. 1(c)]. However, when the local SoPs of the input field are specially tailored, the focusing properties change greatly. For example, when the horizontal index increases form $h = 0$ to $h = 2$, the focal spot firstly gets longer as seen in Fig. 2(b), but then splits to identical twin foci distributions for the electric field in the focal plane as depicted in Figs. 2(c) and 2(d). Further, the distance between the twin foci is tunable controlled by horizontal index $h$. Such high-NA focusing behaviors are much different from the previously reported VOFs. Usually, it is believed that the intensity profiles of the focal field change correspondingly when the input SoPs vary [61-67]. However, there only exhibits a focal shift phenomenon while the intensity profile maintains for large values of horizontal index $h$ of the proposed STLLP beams here.

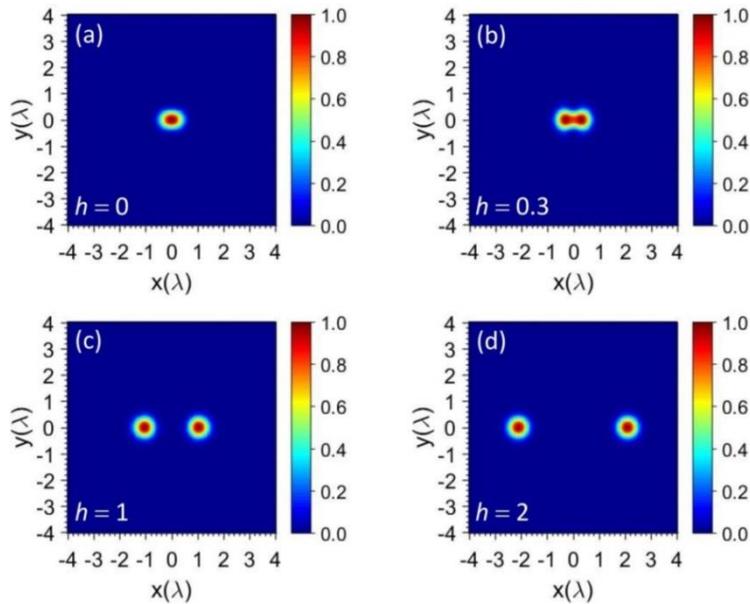

Figure 2: Total electric field intensity distributions in the focal plane of a NA = 0.95 objective lens illuminated with STLLP-LG$_{0,0}$ beams with $h = 0$ (a), 0.3 (b), 1 (c), and 2 (d), respectively.

## 3.2 Partial OAM conversion of optical field from linear state into conjugate OAM states

Based on the extraordinary high-NA focusing behaviors of STLLP-LG$_{0,0}$ beams, it is found that partial OAM conversion of optical field from input linear state into conjugate OAM states takes place, as depicted in Fig. 3, which shows, respectively, detailed focal electric fields of two extreme cases of STLLP-LG$_{0,0}$ beams with $h = 0$ and 2 for comparison. Obviously, the transverse component [Figs. 3(a) and 3(d)] dominates the total field [Figs. 2(a) and 2(d)] for the both cases. Here, we are much interested in their electric fields of longitudinal components. Specially, there are identical twin hot spots with semicircular-shaped patterns when the locally linear SoPs are spatially-invariant [Fig. 3(b)], which exhibit nearly the same phase distributions for each hot spot and a π shift between each other [Fig. 3(c)]. In contrary, they involve to identical twin hollow spot with doughnut-shaped patters when the locally linear SoPS are spatially-tailored [Fig. 3(e)], resulting in opposite helical phase distributions [Fig. 3(f)], which implies the system partially transform linear state into conjugate OAM states [$l = +1$ for the left focus and −1 for the right focus].

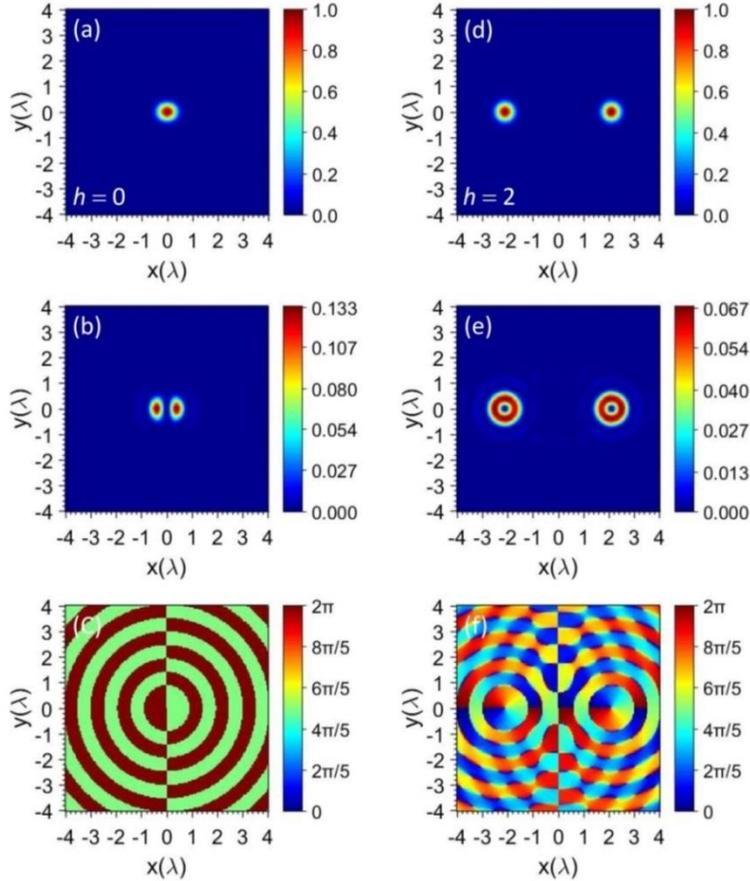

Figure 3: Electric field intensity and phase distributions in the focal plane of tightly focused STLLP-LG$_{0,0}$ beams with $h = 0$ (left column) and 2 (right column), respectively. (a),(d) Intensity distribution of the transverse component. (b),(e) Intensity distribution of the longitudinal component. (c),(f) Phase distribution of the electric field of the longitudinal component.

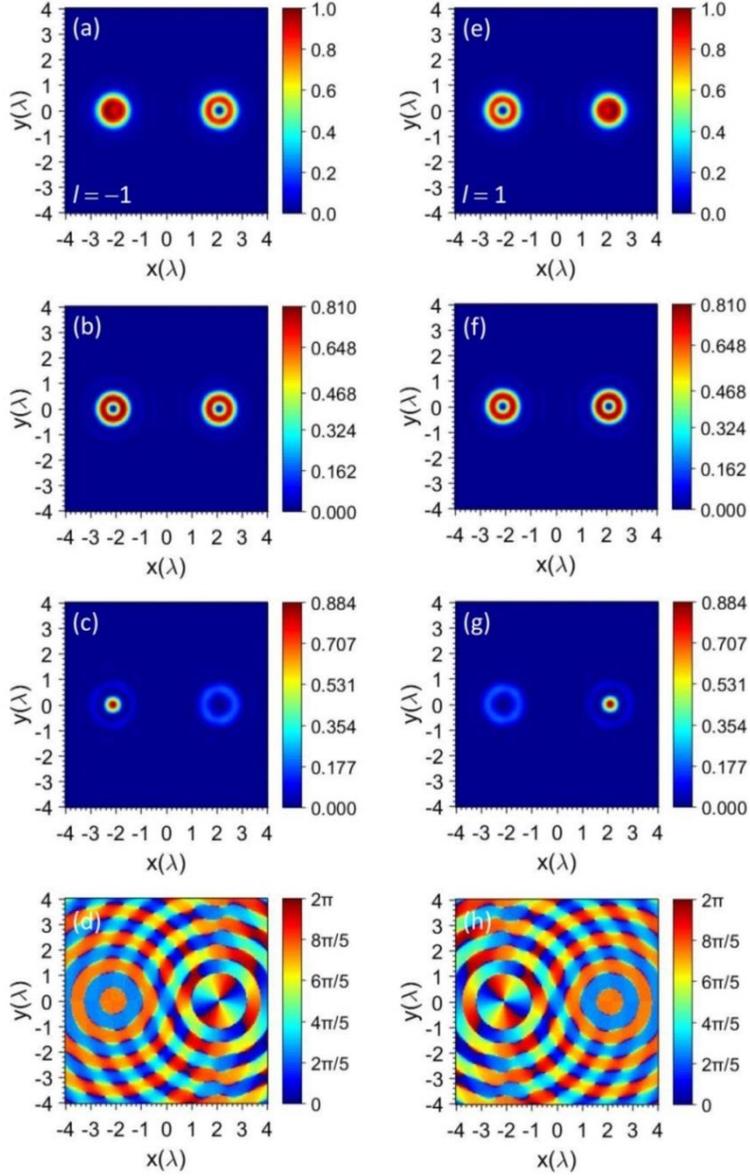

Figure 4: Electric field intensity and phase distributions in the focal plane of tightly focused STLLP-LG$_{-1,0}$ (the left column) and STLLP-LG$_{1,0}$ (the right column) beams when $h = 2$, respectively. (a),(e) Total intensity distribution. (b),(f) Intensity distribution of the transverse component. (c),(g) Intensity distribution of the longitudinal component. (d),(h) Phase distribution of the electric field of the longitudinal component.

To further prove the partial OAM conversion of optical field from linear state into conjugate OAM states, we consider STLLP-LG$_{-1,0}$ and STLLP-LG$_{1,0}$ beams with $h = 2$, which carry none SAM and, respectively, $-\hbar$ and $+\hbar$ OAM per photon. For both cases, the double focal spots are much different from each other for their total fields [Figs. 4(a) and 4(e)], though they are identical for their transverse components [Figs. 4(b) and 4(f)], arriving from the peculiar electric field distributions of longitudinal components [Figs. 4(c) and 4(g)]. To be specific, there is no helical phase related to OAM for the left [right] focus in the longitudinal component, as depicted in Fig. 4d [Fig. 4f], which is caused by the compensation of the opposite OAM arriving from OAM conversion of optical field from linear state into conjugate OAM states. In contrary, double helical phase appears for the right [left] focus shown in Fig. 4d [Fig. 4f], where $\exp[(-1-1)i\varphi] = \exp(-2i\varphi)$ $[\exp[(1+1)i\varphi] = \exp(2i\varphi)]$. These phase changes are manifested in the intensity profiles; they are bright spots for the left [right] focus of the longitudinal component and total fields depicted, respectively, in Figs. 4(c) and 4(a) [Figs. 4(g) and 4(e)], while for the right [left] focus, however, they exhibit donut-shaped patters for longitudinal component and total fields given, respectively, in Figs. 4(c) and 4(a) [Figs. 4(g) and 4(e)].

## 3.3 Energy flow

For giving a deep understanding of partial OAM conversion of optical field from linear state into conjugate OAM states, we now explore the energy flow of the strongly focused STLLP-LG$_{l,p}$ beams. Similarly, when the incident optical field is embodied by Eq. (5), the corresponding magnetic field in the vicinity of focus can be derive as [60]

$$\mathbf{H}(\rho,\phi,z) = \frac{-ikf}{2\pi} \int_0^{2\pi}\int_0^{\alpha} \sin\theta\sqrt{\cos\theta}\,l(\theta)\exp(il\varphi)\mathbf{M_m}e^{ik[-\rho\sin\theta\cos(\varphi-\phi)+z\cos\theta]}\mathrm{d}\varphi\mathrm{d}\theta, \qquad (9)$$

$$\mathbf{M_m} = \begin{bmatrix} \left[-\cos\left(\frac{2\pi h\sin\theta\cos\varphi}{\sin\alpha}-\varphi\right)\sin\varphi - \sin\left(\frac{2\pi h\sin\theta\cos\varphi}{\sin\alpha}-\varphi\right)\cos\theta\cos\varphi\right]\hat{\mathbf{e}}_x \\ \left[\cos\left(\frac{2\pi h\sin\theta\cos\varphi}{\sin\alpha}-\varphi\right)\cos\varphi - \sin\left(\frac{2\pi h\sin\theta\cos\varphi}{\sin\alpha}-\varphi\right)\cos\theta\sin\varphi\right]\hat{\mathbf{e}}_y \\ \left[-\sin\left(\frac{2\pi h\sin\theta\cos\varphi}{\sin\alpha}-\varphi\right)\sin\theta\right]\hat{\mathbf{e}}_z \end{bmatrix}, \qquad (10)$$

where $\mathbf{M_m}$ is the magnetic polarization vector in the strongly focused field. Obviously, the distributions of the magnetic polarization vector described by Eq. 10 are much different from that of the electric polarization vector embodied by Eq. (7). In terms of the full time-dependent three-dimensional electric and magnetic field vectors, the energy current can be defined by the time-averaged Poynting vector as [60]

$$\langle \mathbf{S} \rangle \propto \frac{c}{8\pi} \mathrm{Re}(\mathbf{E} \times \mathbf{H}^*), \tag{11}$$

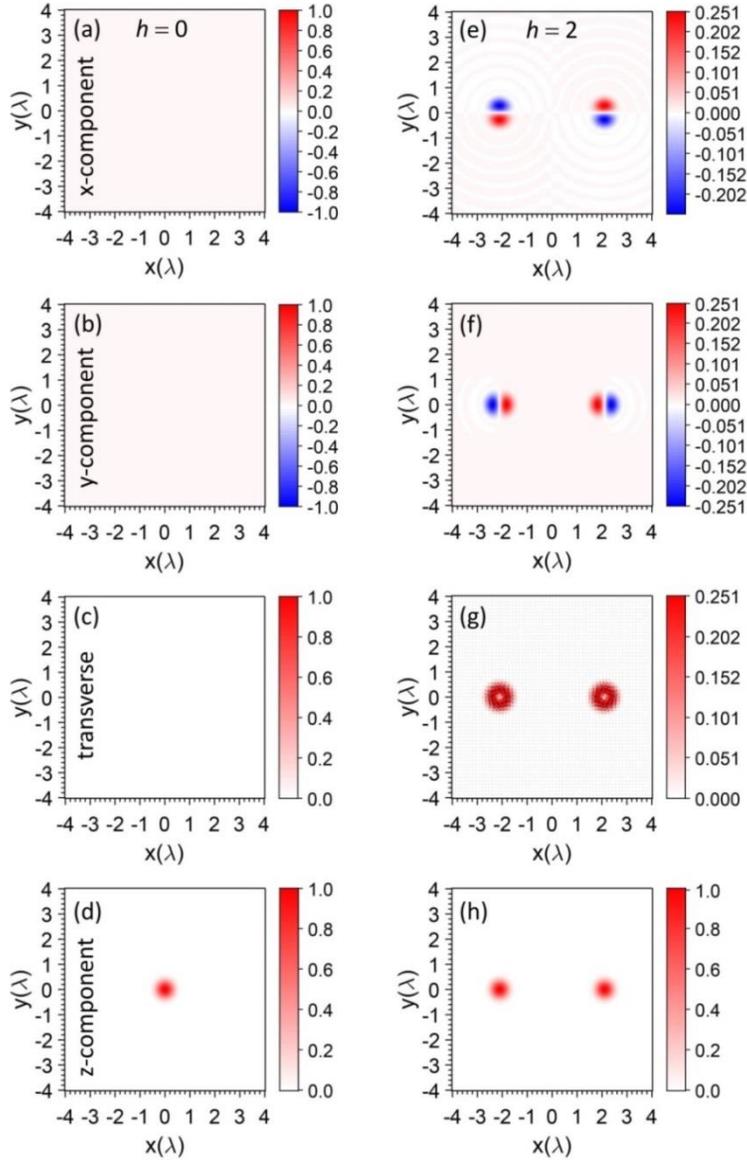

Figure 5: Energy flow in the focal plane of tightly focused STLLP-LG$_{0,0}$ with $h = 0$ and 2, respectively. (a),(e) $x$-component. (b),(f) $y$-component. (c),(g) Transverse component, the direction of which is shown by the black arrows. (d),(h) $z$-component.

where the **E** and **H** are the electric and magnetic fields in the image space, respectively, asterisk denotes the operation of complex conjugation. Then, we can calculate the energy flow of strongly focused STLLP-LG$_{l,p}$ beams with Eqs. (6)-(11). The focusing conditions are the same as that mentioned above.

For the sake of comparison, figure 5 gives the energy flow in the focal plane of two extreme cases of STLLP-LG$_{0,0}$ beams with $h = 0$ and 2, respectively. Traditionally, it is believed that there is no and ring-shaped transverse energy flow in the focal plane of strongly focused LLP plane and vortex beams [68-71], respectively. It is true for the LP plane beam shown in the left column in Fig. 5, wherein only the $z$-component Poynting vector appears and exhibits a bright spot distribution located on axis. However, double energy flow rings emerge with opposite rotational directions for the transverse component of the proposed STLLP-LG$_{0,0}$ plane beam when $h = 2$, which further proves that the system successfully achieve partial OAM conversion from linear state to conjugate OAM states in in homogeneous and isotropic medium. The corresponding $z$-component Poynting vector is shown in Fig. 5(h), the properties that identical twin bright spots play the dominate role are very obvious.

## 4 Conclusion

To summarize, we have demonstrated that OAM conversion of light without any spin state is possible in a homogenous and isotropic medium, provided the STLLP beam is strongly focused. In such a strong focusing system, STTLPs can generate identical twin foci with tunable distances between them controlled by the input SoPs for all the transverse, longitudinal, and total fields in the focal region. Most importantly, such system achieves partial OAM conversion from linear state to conjugate OAM states, giving rise to helical phases with opposite directions for each focus in the longitudinal component of the electric field. Due to such OAM conversion, there exhibits twin transverse energy flow rings with opposite directions in the focal plane. The possibility of interconverting optical linear and OAM states may offer new fundamental insights into the nature of light and facilitate the development of new generations of nanoscale optical manipulation techniques.

This work was partially supported by National Natural Science Foundation of China (NSFC) (11604182, 11704226); Natural Science Foundation of Shandong Province (ZR2016AB05, ZR2017MA051); Key Laboratory of Optoelectronic Devices and Systems of Ministry of Education and Guangdong Province (GD201704).